\documentclass[12pt]{article}

\usepackage{siunitx}
\usepackage{mathrsfs}  
\usepackage{units}
\usepackage{bbm}
\usepackage{multicol}

\usepackage{graphicx}
\usepackage{tabularx}
\usepackage{a4}
\usepackage{url}
\usepackage{longtable, lscape}
\usepackage[usenames, dvipsnames]{color}
 \usepackage{pdflscape}

\usepackage{multirow, makecell, bigstrut, booktabs, float}
\usepackage{array}

\usepackage{setspace}
\usepackage[fleqn]{amsmath}
\usepackage{amsfonts}
\usepackage{amssymb}
\usepackage{subfigure}
\usepackage{tcolorbox}
\usepackage{mathrsfs}  
\usepackage{bbm}

\usepackage{adjustbox}
\usepackage{nicefrac}
\usepackage{pgfplots}
\usepackage{authblk}
\usepackage[fleqn]{amsmath}
\usepackage{float}
\usepackage[section]{placeins}
\usepackage{subfigure}
\usepackage{pgfplots}
\graphicspath{{Bilder/}}

\begin{document}

\title{The population-attributable fraction for time-dependent exposures using dynamic prediction and landmarking}

\author[1,2]{Maja von Cube\thanks{cube@imbi.uni-freiburg.de, Ernst-Zermelo Strasse 1, 79104 Freiburg}}

\author[1,2]{Martin Schumacher}

\author[3]{Hein Putter}
\author[4,5]{J\'{e}an-Fran\c{c}ois Timsit}

\author[6]{Cornelis van de Velde}
\author[1,2]{Martin Wolkewitz}

\affil[1]{Institute of Medical Biometry and Statistics, Faculty of Medicine and Medical Center - University of Freiburg, Freiburg, Germany}

\affil[2]{Freiburg Center for Data Analysis and Modelling, University of Freiburg, Freiburg, Germany}

\affil[3]{Leiden University Medical Centre, Department of Medical Statistics and Bioinformatics}

\affil[4]{UMR 1137 IAME Inserm/Universit\'{e} Paris Diderot, Paris, France}
\affil[5]{APHP Medical and Infectious Diseases ICU, Bichat Hospital, Paris, France}

\affil[6]{Leiden University Medical Centre, Department of Surgery, Leiden, Netherlands}

\date{April 5, 2019}

\renewcommand\Authands{ and }

\maketitle


\maketitle                   
\newpage
\begin{abstract}
The public health impact of a harmful exposure can be quantified by the population-attributable fraction (PAF). The PAF describes the attributable risk due to an exposure and is often interpreted as the proportion of preventable cases if the exposure could be extinct.

Difficulties in the definition and interpretation of the PAF arise when the exposure of interest depends on time. Then, the definition of exposed and unexposed individuals is not straightforward.

We propose dynamic prediction and landmarking to define and estimate a PAF in this data situation. Two estimands are discussed which are based on two hypothetical interventions that could prevent the exposure in different ways.

Considering the first estimand, at each landmark the estimation problem is reduced to a time-independent setting. Then, estimation is simply performed by using a generalized-linear model accounting for the current exposure state and further (time-varying) covariates. The second estimand is based on counterfactual outcomes, estimation can be performed using pseudo-values or inverse-probability weights.

The approach is explored in a simulation study and applied on two data examples.
First, we study a large French database of intensive care unit patients to estimate the population-benefit of a pathogen-specific intervention that could prevent ventilator-associated pneumonia caused by the pathogen \textit{Pseudomonas aeruginosa}. Moreover, we quantify the population-attributable burden of locoregional and distant recurrence in breast cancer patients.

\textbf{keywords: }Attributable risk; Competing risks; Dynamic prediction; Landmarking; Time-dependent exposure;

\end{abstract}

\section{Introduction}

A critical aspect of understanding the consequences of an exposure is the harm it causes for the entire population under study.
Whereas the relative risk (RR), and the odds ratio (OR) are commonly used to quantify the risk increase at the individual level, the population-attributable fraction (PAF) quantifies attributable risk at the population level. 
The PAF is defined as the proportion of attributable cases due to exposure and is often interpreted as the proportion of preventable cases if the exposure were extinct. 

The PAF was initially defined for basic study designs such as cohort studies of fixed length (\cite{Levin1953PAF}). However, data situations are often more complex. For example, in hospital epidemiology the goal may be a quantification of the burden of hospital-acquired infections (HAIs). As HAIs occur over the course of hospital stays, patients are naturally unexposed at time of admission to the hospital. The time-dependency of exposure makes it difficult to define the group of patients who are considered to be unexposed. For example, treating eventually exposed patients as exposed since study entry results in the immortal time bias (\cite{schumacher2013hospital}). Moreover, as the outcome of interest is often death in the hospital, discharge alive must be considered as a competing event (\cite{wolkewitz2014interpreting}). Finally, adjustment for confounding is essential to draw causal conclusions from observational data. Since the kind of exposure we consider depends on time, we must also consider time-varying confounding. Corresponding adjustments appear challenging due to collider-stratification bias (\cite{greenland2003quantifying}).

The concept of the PAF has been extended to accommodate time-to-event data with an exposure that is fixed at baseline (\cite{chen2006attributable, chen2010attributable, samuelsen2008attributable, sjolander2016cautionary, zhao2017onestimation}). Extensions of the PAF to data settings with a binary time-dependent exposure and competing risks have been proposed by \cite{schumacher2007attributable} and \cite{bekaert2010adjusting}. Both proposed estimands are cumulative measures of attributable risk over the course of time. Thus, they provide information on the evolution of the PAF for the \textit{complete} target population. As a consequence, the effect of a preventive intervention for the subgroup of individuals who are still at risk at later time points may be precluded by the cumulative nature of the estimands.

In this article, we propose two novel estimands of the PAF that account for a \textit{time dynamic} target population. Being based on dynamic prediction and landmarking (\cite{van2011dynamic}), the estimands allow for a differentiation between individuals with differing durations at risk. These patients may differ strongly due to evolving patient characteristics. Consider, for example, a population of ICU patients. Patients being a long time at risk  -- so-called long-stayers -- are often sicker and also longer at risk to acquire the exposure (e.g. an HAI).

The first proposed estimand is based on a hypothetical intervention that is only effective at the actual time of intervention. The estimand is defined over a specific time window within the study time scale. It summarizes the effect of intervention within this specific time frame rather than over the complete follow-up time. A fixed time window is less dependent on the arbitrary end of follow-up time point, which depends on the patients in the study. The second estimand is based on an intervention that is effective for a certain amount of time. Thus, this estimand allows for more realistic long-term intervention relaxing the assumptions of the first estimand. Moreover, it is also defined over a specific time window.

After an introduction of the methods, we present the results of a simulation study in which we investigate the behaviour of the estimands and estimators.
Subsequently, the approaches are applied to a real data sample of ventilated patients in intensive care. This patient population is at high risk of acquiring ventilator-associated pneumonia (VAP) \cite{chastre2002ventilator}). VAPs are predominantly caused by the pathogen \textit{Pseudomonas aeruginosa} (\textit{Pa}). To understand the potential benefit of pathogen-specific intervention programs, we study the percentage of preventable deaths of ventilated patients in the ICU if VAP caused by \textit{Pa} could have been avoided. 

The method devised in this article is motivated by a particular data example. However, it is applicable to any other data setting with binary time-dependent exposure. Therefore, we also apply the method within a different data context. Based on the study of \cite{fontein2015dynamic} on breast cancer in the TEAM trial data, we estimate the proportion of attributable death cases due to distant and locoregional recurrence. The data example serves as a demonstration of alternative adjustment methods and is an application in basic survival settings with internal time-dependent exposures but no competing risks.
The article ends with a discussion.

\section{The population-attributable fraction for cohort studies of fixed length with a baseline exposure}

The PAF has been defined by \cite{benichou2001review} as
\begin{equation}
PAF=\frac{P(D=1)-P(D=1|E=0)}{P(D=1)},
\label{PAFl}
\end{equation}
where $D$ is the random variable of a dichotomous outcome and $E$ of a dichotomous exposure. The realization of both $D$ and $E$ is observable. The proposed estimand is valid for cross-sectional studies and cohort studies of fixed length with a baseline exposure.

The PAF is usually estimated with observational data, as the effect of the exposure is generally considered to be harmful. Therefore, adjustment for confounding is essential to obtain an unbiased estimate of the population-attributable burden.

\cite{miettinen1974proportion} showed that with $P(D=1)=P(E=1)P(D=1|E=1)+P(E=0)P(D=1|E=0)$ an equivalent definition of the PAF is
\begin{equation}
PAF=P(E=1|D=1)\times \frac{RR-1}{RR},
\label{PAFm}
\end{equation} 
where
\begin{equation*}
RR=\frac{P(D=1|E=1)}{P(D=1|E=0)}.
\end{equation*} 
This reformulation of the PAF allows for a straightforward way of adjusting the estimator of the PAF by plug-in of an adjusted RR in definition \eqref{PAFm} (\cite{miettinen1974proportion}). The prevalence of the exposure among cases, $P(E=1|D=1)$, can be estimated by the population average without further consideration of confounders (\cite{miettinen1974proportion}).

\section{Conventional dynamic prediction and landmarking to estimate the population-attributable fraction}

Dynamic prediction and landmarking implies that estimands are defined for a set of time points (landmarks) during follow-up. This allows for an update of the target population that is still at risk at the specific landmark (LM). At each LM only those individuals at risk are considered, and their exposure state and other patient characteristics are updated at each LM. Thus, at each LM a different target population is being considered.

A major strength of conventional dynamic prediction and landmarking is the facility of adjustment for confounding of the estimator. At each LM the data situation is reduced to a time-independent setting. The resulting estimand differs from definitions \eqref{PAFl} and \eqref{PAFm} only by the target population. Thus, adjustment for confounding is based on established methods available for estimands that ignore time of exposure and outcome, such as the approach proposed by \cite{miettinen1974proportion}.

\subsection{Formal definition of $PAF_{LM,h}$}

The estimand $PAF_{LM,h}$ consists of a set of separate estimands $PAF(l,h)$. These depend on a specific landmark $l$ and a time window $h$ and are formally given by 
\begin{equation}
PAF(l,h)=\frac{P(D_{l,h}=1|A_l=1)-P(D_{l,h}=1|E_l=0,A_l=1)}{P(D_{l,h}=1|A_l=1)},
\label{PAF(s,t)_lev}
\end{equation}
where $A_l$ is the random variable denoting the patients' at-risk state at $l$, $E_l$ is the random variable of the exposure state at $l$, and $D_{l,h}$ is the random variable of the occurrence of the outcome within $(l,l+h]$.

Since $P(D_{l,h}=1|A_l=1)=P(E_l=1|A_l=1)P(D_{l,h}=1|E_l=1, A_l=1)+P(E_l=0|A_l=1)P(D_{l,h}=1|E_l=0, A_l=1)$, an equivalent definition of $PAF(l,h)$ is
\begin{align}
PAF(l,h)&=P(E_l=1|A_l=1, D_{l,h}=1)\times \frac{RR_{l,h}-1}{RR_{l,h}}\nonumber\\
 &=P_{E_l}\times \frac{RR_{l,h}-1}{RR_{l,h}}
 \label{PAF(s,t)_def}
\end{align}
with $P_{E_l}$ being the prevalence of the exposure at time $l$ among cases occurring within the time window $(l,l+h]$ and $RR_{l,h}$ is the RR of the outcome within $(l,l+h]$ depending on the exposure state at time point $l$, i.e.
\begin{equation}
RR_{l,h}=\frac{P(D_{l,h}=1|A_l=1,E_l=1)}{P(D_{l,h}=1|A_l=1, E_l=0)}.
\end{equation}

Formally, we define by $PAF_{LM,h}$ the set of the $PAF(l,h)$s over all LMs, i.e.
$PAF_{LM,h}=\{PAF(l,h)|l\in \mathcal{LM}\}$, where  $\mathcal{LM}$ is the set of chosen LMs and $h$ is the length of the time window.

\subsection{Estimation of $PAF(l,h)$}

For estimation, the LMs should be chosen based on the number of exposed and unexposed patients at each time point. The number in both patient groups should be sufficiently large for inference. The time window of fixed length can be defined based on clinical knowledge and relevance. Examples of how to choose the LMs and time windows are presented in Sections 5 to 7 based on a simulation study and real data examples.

Inference at a LM $l$ is performed on a LM dataset (\cite{van2011dynamic}) which consists of all patients who are still at risk at the LM. For every patient within the LM dataset the exposure state at the LM as well as the event state at the end of the time window is accessed. 
Patients who do not experience the event of interest within the time window are implicitly assumed to be either still at risk at the end of the time window or to have experienced the competing event. 
The matter of censoring is discussed at the end of this section. 

Using \eqref{PAF(s,t)_def} to define $PAF(l,h)$ results in a straightforward approach of adjustment for confounding (see \cite{miettinen1974proportion}). As explained by \cite{miettinen1974proportion} and briefly described in Section 2, $PAF(l,h)$ can be estimated by separately estimating the prevalence $P_{E_l}$ and $RR_{l,h}$. An estimator that accounts for systematic differences between exposed and unexposed patients is obtained by adjusting $RR_{l,h}$ at each LM $l$.

Estimation of the adjusted $RR_{l,h}$ can be performed via maximum likelihood estimation using a $\log$-model, i.e.
\begin{equation}
P(D_{l,h}=1|A_l=1, E_l, Z_l)=\exp(\beta_{0l}+\beta_{1l}\times E_l+\beta_l^T\times Z_l),
\end{equation}
where $\beta_{0l}$ is the intercept at LM $l$, $\beta_{1l}$ is the coefficient of the exposure state at $l$, $\beta_l^T$ is the vector of coefficients of $Z_l$.
$Z_l$ is a vector of baseline and time-dependent covariates (with covariate values fixed at time point $l$) sufficient for confounder control. The condition $A_l=1$ is automatically fulfilled, as estimation is based on the LM dataset. 
The $RR_{l,h}$ is estimated by $\exp(\widehat{\beta}_{1l})$. The $\log$-model is implemented in the statistical software R within the function \textit{glm} (\cite{stats2017}).

The prevalence $P_{E_l}$ is simply estimated by the proportion of exposed subjects among cases. With \textit{glm}, $P_{E_l}$ could, for example, be estimated by
\begin{equation}
P_{E_l}=P(E_l=1|A_l=1, D_{l,h})=\exp(\alpha_{0l}+\alpha_{1l}\times D_{l,h}),
\end{equation}
where $\alpha_{0l}$ is the intercept at LM $l$ and $\alpha_{1l}$ the regression coefficient of $D_{l,h}$.
An estimator of the variance fo $\widehat{PAF}(l,h)$ is obtained analogously to the variance estimator of the time-fixed PAF \eqref{PAFm} (for details see \cite{greenland1987variance}).

\subsubsection{Smoothing methods for the separate $PAF(l,h)$}

For practical reasons \cite{van2011dynamic} propose to smooth the separate estimates of $PAF(l,h)$ over all LMs. The advantages of smoothing are not only the removal of noise and the nicer presentation, but also the availability of values of $PAF_{LM,h}$ between two LM time points. Moreover, depending on the smoothing method used, an increase in efficiency can be achieved. 

In principle, two approaches can be used to obtain a smooth curve that interpolates the separate quantities $PAF(l,h)$ over all LMs $l$, $l\in \mathcal{LM}$. 
An ad hoc approach is to first estimate $PAF(l,h)$ for each LM separately (as described in Section 3.2). Then, smoothing splines, regression splines, polynomial regression or local regression can be used to fit a smooth curve on the separate estimates. The methods are well described by \cite{james2013introduction}. The downside of this approach is that the curve may be easily overfitted, as the choice of the degree of smoothing is not based on any statistical tests. Therefore, the gain in efficiency is negligible.

As an alternative the principle of pooled logistic regression by a so-called supermodel can be used to obtain a smooth curve over all LMs directly without first fitting the separate models (\cite{van2011dynamic}). Estimation with the supermodel is based on the LM datasets stacked together to one large dataset. In this "super prediction dataset" the patients are represented as often as they appear in the LM datasets (see \cite{van2011dynamic}). A regression model is fitted on the super prediction dataset separately for $RR_{l,h}$ and $P_{E_l}$ by accounting for possible time-varying effects at the LMs via interaction terms. 

A supermodel for $P_{E_l}$ is given by
\begin{equation}
P_{E_l}=P(E_l=1|A_l=1, D_{l,h})=\exp(\alpha_0(l)+\alpha_1(l) \times D_{l,h})
\end{equation}
where $\alpha_0(l)=\sum \alpha_ {00} \times f_0(l)+...+ \alpha_{0K} \times f_K(l)$ and $\alpha_1(l)=\sum \alpha_ {10} \times g_0(l)+...+ \alpha_{1\tilde{K}} \times g_{\tilde{K}}(l)$, with $f_0(l)=g_0(l)=1$ and $f_j(l)$, $g_j(l)$ ($1<j\leq K \text{ or resp. } \tilde{K}$) being smooth basis functions of the LMs $l$, $l\in \mathcal{LM}$. We choose our basis functions as proposed in most papers about dynamic prediction and landmarking (\cite{van2008dynamic,  nicolaie2013CR, nicolaie2013dynamic}). Thus, $K=\tilde{K}=2$ and $f_1(l)=g_1(l)=l$, $f_2(l)=g_2(l)=l^2$. This results in the following supermodel
\begin{equation}
P_{E_l}=\exp(\alpha_{00}+\alpha_{01}l+\alpha_{02}l^2+\alpha_{10}\times D_{l,h}+\alpha_{11}l\times D_{l,h}+\alpha_{12}l^2\times D_{l,h}).
\end{equation}
We use the same basis functions to estimate $RR_{l,h}$:
\begin{equation}
P(D_{l,h}=1|A_l=1, E_l, Z_l, l)=\exp(\beta_0(l)+\beta_1(l)\times E_l+\beta^T\times Z_l),
\end{equation}
with $\beta_0(l)=(\beta_{00}+\beta_{01}l+\beta_{02}l^2)$ and $\beta_1(l)=(\beta_{10}+\beta_{12}l+\beta_{12}l^2)$. For simplicity we do not include any time-varying effects for the covariate vector $Z_l$. $RR_{l,h}$ is now estimated by $\exp(\widehat{\beta_1}(l))$, where $l$ is any time point between the first and the last LM.
The intercept function $\beta_0(l)$ models the baseline risk of experiencing the event of interest within the time interval, given alive and at risk at the LM. When including an interaction term for the intercept, we assume that the risk of experiencing the event of interest within the time interval changes with time. The effect of acquisition of exposure is modelled by the regression coefficients $\beta_{1}(l)$. If interaction terms with the LMs are included, we assume a change of the effect of the exposure with time. 
Whether interaction terms with the LMs are needed and up to which degree can be tested with the Wald test (\cite{van2011dynamic}). Finally, a smoothed estimate of $PAF_{LM,h}$ is obtained by plugging the estimates of $P_{E_l}$ and $RR_{l,h}$ obtained with the supermodel into equation \eqref{PAF(s,t)_def}. For both discussed smoothing approaches, we propose to obtain confidence intervals (CIs) by a bootstrap approach. 

\subsubsection{Dealing with censored observations}

The method explained so far is valid in the presence of competing risks but without censoring. Then, the outcome $D_{l,h}$ is observable for all patients. In hospital epidemiology, this is a common data set up. To generalize the estimation approach to data settings subject to censoring, dynamic pseudo-observations proposed by \cite{nicolaie2013dynamic} can be used. After the pseudo-values have been obtained, the same method as described above (i.e. the approach by \cite{miettinen1974proportion}) can be applied. 

$D_{l,h}$ can be also expressed in terms of a survival probability \cite{nicolaie2013dynamic}. Let $J$ denote the number of different event types and let $X$ indicate the type of the event. Without loss of generality, we assume $J=2$ and $X=1$ for the event of interest. 
We have $D_{l,h}=I(T\leq l+h, X=1|T>l)$, where $T$ is the minimum of the event and censoring time.
A pseudo-value for $D_{l,h}$ is obtained within the LM dataset at $l$. For an individual $i$ at risk at LM $l$ the pseudo-observation is defined as
\begin{equation}
\hat{\theta}^{i}_{l,h}=n_l\hat{F}_1(l+h|l)-(n_l-1)\hat{F}^{(-i)}_1(l+h|l),
\end{equation}
where $n_l$ is the number of patients at risk at $l$, $\hat{F}_1(l+h|l)$ is the Aalen-Johansen estimator for the conditional cumulative incidence function $P(T\leq l+h, X=1|T>l)$, and $\hat{F}^{(-i)}(l+h|l)$ is the according Aalen-Johansen estimator based on the LM data set at $l$ leaving out observation $i$. As suggested by \cite{nicolaie2013CR} the pseudo-observations can be easily obtained with the R package \textit{pseudo} (\cite{pseudo2017}), or alternatively 
with the R package \textit{prodlim} (\cite{prodlim2017}).

Given the LMs satisfy that the time window $(l,l+h]$ does not exceed the study horizon $\tau$ with both the probability of censoring and the probability of still being at risk exceeding zero beyond $\tau$, \cite{nicolaie2013dynamic} derive asymptotic properties for the dynamic pseudo-observations. With these properties, they show that estimation of $P_{E_l}$ and $RR_{l,h}$ can be performed in the same way explained above for both the separate models and the supermodel based on the pseudo-observations. This also applies to the variance estimators. Then, an estimate of $PAF(l,h)$ and the variance can be obtained by plug-in of the estimators of $RR_{l,h}$ and $P_{E_l}$.

An alternative way to handle censored observation is the use of standard survival analysis methodology. This approach is based on definition \eqref{PAF(s,t)_def}. The marginal risk $P(D_{l,h}=1|A_l=1)$ and the conditional risk $P(D_{l,h}=1|A_l=1, E_l=0)$ can both be estimated with the Aalen-Johansen estimator. Adjustment is possible, for example, with the Cox proportional hazards model. This approach is explained in detail by \cite{van2008dynamic}. We also refer to Section 7 where we apply this approach to a specific data example. 

\section{A counterfactual approach to define the population-attributable fraction}

\begin{figure*}[hbtp!]
\centering
\includegraphics[scale=0.6]{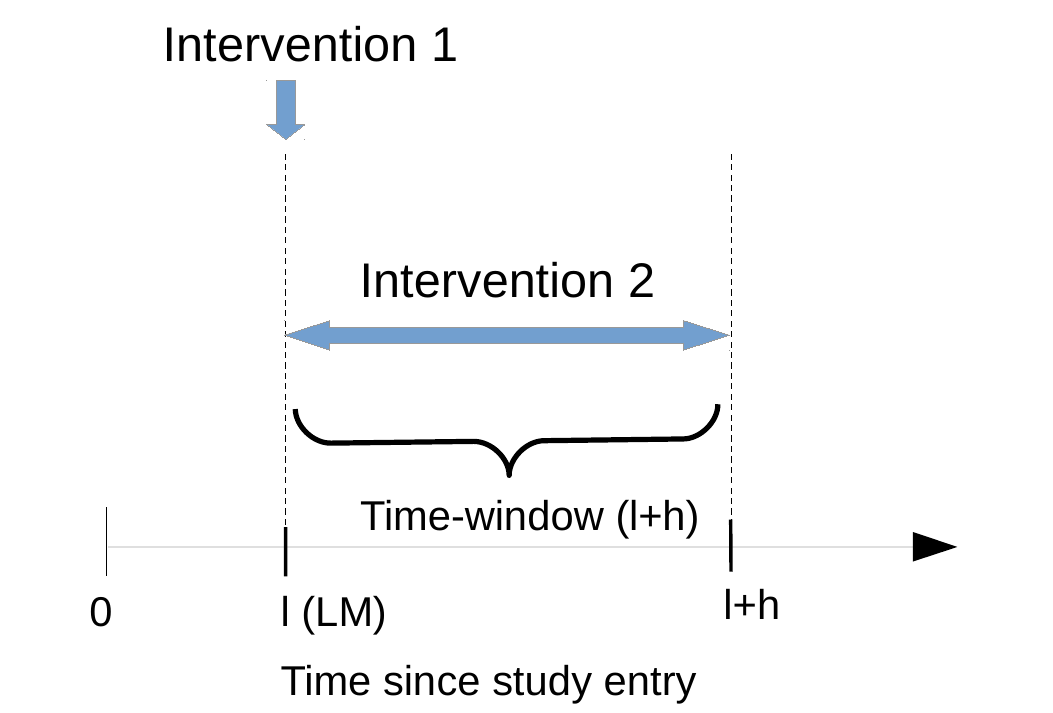}
\caption{Illustration of the two different types of interventions. Intervention 1 corresponds to the estimand defined by conventional dynamic prediction and landmarking, $PAF_{LM,h}$, intervention 2 corresponds to the estimand defined by its' extension, $PAF_{0 LM,h}$.}
\label{fig:LMinter}
\end{figure*}

The estimand $PAF_{LM,h}$ is interpretable as the percentage of preventable cases if the exposure could be eliminated \textit{at} the LM.
However, interventions often have a long-term effect. Thus, a possibly more desirable interpretation could be 'the proportion of preventable cases if the exposure could be avoided over the entire time interval $[l,l+h]$'. We denote the estimand with this interpretation by $PAF_0(l,h)$ and the set of estimands over a range of LMs $l$ by $PAF_{0 LM,h}$.
A graphical illustration of the interventions assumed for $PAF_{LM,h}$ and $PAF_{0 LM,h}$ is given in Figure \ref{fig:LMinter}.

\subsection{Formal definition of $PAF_{0LM,h}$}

Formally, $PAF_{0}(l,h)$ can be defined as
\begin{equation}
PAF_{0}(l,h)=\frac{P(D_{l,h}=1|A_l=1)-P(D_{0 l,h}=1|A_l=1)}{P(D_{l,h}=1|A_l=1)}.
\label{PAF(s,t)_cens}
\end{equation} 
$P(D_{l,h}=1|A_l=1)$ is the (observable) overall risk of experiencing the event of interest within the time window $(l,l+h]$ among patients still alive at $l$. $P(D_{0 l,h}=1|A_l=1)$ is the hypothetical risk of experiencing the event of interest within $(l,l+h]$ among patients still alive at $l$ had the exposure been removed within the time window $[l,l+h]$ for all patients who were still at risk at $l$.

We expressed $PAF_0(l,h)$ with marginal outcome risks. This definition corresponds to definition \eqref{PAF(s,t)_lev} of $PAF(l,h)$. $PAF(l,h)$ was also expressed in terms of RR and prevalence, which is defined by the conditional outcome risks.
The two definitions \eqref{PAF(s,t)_lev} and \eqref{PAF(s,t)_def} are equivalent. This is due to the fact that the exposure state is fixed at LM $l$. In contrast, $PAF_0(l,h)$ cannot be reformulated in terms of RR and prevalence, because the reformulation uses information about the prevalence within the time window which is unknown at $l$. Formally, we would need $P(D_{l,h}=1|A_l=1, E_l=0)=P(D_{0 l,h}=1|A_l=1)$, which is not the case if some patients acquire the exposure within the window $(l,l+h]$ (post time point $l$).

\subsection{Estimation of $PAF_0(l,h)$}

To estimate $PAF_0(l,h)$, we separately estimate $P(D_{l,h}=1|A_l=1)$ and $P(D_{0 l,h}=1|A_l=1)$ and plug them into equation \eqref{PAF(s,t)_cens}. Estimation of the observable marginal risk $P(D_{l,h}=1|A_l=1)$ is straightforward by maximum likelihood estimation based on generalized linear models (glms). A possible model is
\begin{equation}
P(D_{l,h}=1|A_l=1)=\text{expit}(\beta_{0l}).
\end{equation}
In this situation other link functions such as the link function $\exp$ would return identical results.

Estimation of the hypothetical risk of the outcome had all patients been unexposed, $P(D_{0 l,h}=1|A_l=1)$, is complicated by the missing values of exposed patients. Given consistency the counterfactual random variable of the event of interest had the exposure been removed, $D_{0 l,h}$, is only realized for patients who were factually observed to remain unexposed within $[l,l+h]$. For patients who acquire the exposure within the time window the realization of $D_{0 l,h}$ is not observable. Therefore, it must be imputed based on the information available for patients who were observed to remain unexposed within $[l,l+h]$.

To impute the missing outcomes of exposed patients various approaches are possible. In the following, we first explain how to use a pseudo-value approach similar to Section 3 where this approach was used for the imputation of missing outcomes due to censoring. 
With this approach, adjustment for time-dependent covariates updated at each LM is possible. However, in contrast to the estimation of $PAF_{LM,h}$, which assumes an intervention that is effective at the LM only, unbiased estimation of $PAF_{0 LM,h}$ requires patients to be comparable within the complete time window $h$ and not just at the LM $l$. Therefore, adjustment for time-varying confounding within the time window might become necessary. Adjustment for time-varying confounding can be done using inverse-probability of exposure weights. This procedure is also explained below.

We first explain the pseudo-value approach for an estimation of $P(D_{0 l,h}=1|A_l=1)$. In principle, we do the same as in Section 3.2.3. However, instead of dealing with outcomes that are missing due to censoring, we aim to impute the outcomes of patients who acquired the exposure within the time window. The crucial assumption is that patients who did not acquire the exposure within the time window are comparable to those patients who did acquire it. 

An estimator of $P(D_{0 l,h}=1|A_l=1)$ can be obtained by treating patients who acquire the exposure within the time window as censored. 
We express the realization of $D_{0 l,h}$ by $I(T_0\leq l+h, X=1|T>l)$, where $T_0$ is the minimum of the event and exposure time and $X$ is, as in Section 3.2.3., the event indicator (i.e. $X=1$ if the patient experienced the event of interest in $(l, l+h]$, $X=2$ if the patient experienced the competing event). Then, the pseudo-value of $D_{0 l,h}$ for patient $i$ is defined as
\begin{equation}
\hat{\theta}^{i}_{0 l,h}=n^0_l\hat{F}^0_1(l+h|l)-(n^0_l-1)\hat{F}^{0(-i)}_1(l+h|l),
\end{equation}
where $n^0_l$ is the number of patients being unexposed and at risk at time point $l$, $\hat{F}^0_1(l+h|l)$ is the Aalen-Johansen estimator for the conditional cumulative incidence function $P(T_0\leq l+h, X=1|T>l)$ and $\hat{F}^{0(-i)}(l+h|l)$ is the according Aalen-Johansen estimator based on the LM data set at $l$ leaving out observation $i$. In contrast to the pseudo-values in Section 3.2.3., $\hat{\theta}^{i}_{0 l,h}$ is estimated with the patients that are unexposed at LM $l$. In Section 3.2.3., pseudo-values were estimated with all patients at risk at $l$. This is because we address two distinct problems. In Section 3.2.3. the \textit{observable} realization of the random variable $D_{l,h}$ was missing due to censoring. Here, we aim to estimate pseudo-values for $D_{0 l,h}$ by
assuming that patients who acquire the exposure within the time window $(l,l+h]$ would have the same risk to experience an event as patients who remain unexposed. Thus, patients who acquire the exposure must be treated as censored.
After the pseudo-values have been obtained $P(D_{0 l,h}=1|A_l=1)$ can be also estimated with glms using the R-package \textit{geese}.

The estimate hereby obtained is most likely biased due to a systematic difference between patients who acquire the exposure and patients who remain unexposed. Such confounding implies a violation of the assumption that patients exposed at some time point within $[l,l+h]$ would have the same risk as patients remaining unexposed within $[l,l+h]$.

As discussed by \cite{andersen2010pseudo} and \cite{binder2014pseudo}, the pseudo-observations are based on the assumption of covariate independent censoring. In the presence of confounding this assumption is violated. If adjustment for covariates updated at each LM is sufficient for confounder control, adjustment of $P(D_{0 l,h}=1|A_l=1)$ can be done by using covariate-adjusted pseudo-values for the estimation with \textit{geese}. The adjusted pseudo-values are based on the adjusted Aalen-Johansen estimators $\hat{F}^0_1(l+h|l,E_l=0, Z_l)$, where $Z_l$ is a vector of covariate values observed at LM $l$.
Yet up to now, the R functions \textit{pseudo} (\cite{pseudo2017}) and \textit{prodlim} (\cite{prodlim2017}) are not implemented for covariate adjusted Aalen-Johansen estimators. Therefore, the pseudo-values must be adjusted either via stratification or - more sophisticated - via nearest-neighbour estimation as proposed by \cite{beran1981nonparametric}.

However, with increasing length of the time window $h$, adjustment for time-varying confounding becomes necessary. 
Therefore, we propose to use adjusted inverse probability of exposure weights as initially proposed by \cite{hernan2000marginal}. The approach is similar to the modified sequential Cox approach proposed by \cite{karim2018comparison}.
The weights are estimated separately for each LM dataset using the R package \textit{ipw}. Then, based on the LM datasets, two weighted cause-specific Cox regression models -- one for the occurrence of the event of interest within the time window and one for the occurrence of the competing event within the time window -- are used to estimate the effect of the exposure within the time window. This can be done with the R function \textit{coxph} of the \textit{survival} package (\cite{survival2015}). Patients who remain at risk until $l+h$ are administratively censored at the end of the time window. Finally, $P(D_{0 l,h}=1|A_l=1)$ can be estimated using the two Cox regression models and the R function \textit{survfit}. A detailed description of this procedure is provided in the tutorial by \cite{Therneau2016multi} and the Appendix of \cite{pouwels2018intensive}. To obtain appropriate CIs, we propose to use a bootstrap method. Smoothed estimates can be obtained as explained in Section 3.

\section{Simulation study}

In a simulation study, we investigated the performance of the two proposed estimands and their estimators. It serves as an exemplary presentation of the two approaches under controllable conditions. In the following, we present two simulation scenarios. Other simulations based on different models have been performed. They led to the same conclusions and are not presented here. 

The two simulation scenarios were based on a multi-state model considering the binary time-dependent exposure as an intermediate event. The final absorbing states were the event of interest and a competing event (see Figure S1 in the online supplementary material). The resulting model is often called extended illness-death model. A detailed description of the model is given by \cite{beyersmann2011competing}.
Both estimands, $PAF_{LM,h}$ and $PAF_{0LM,h}$, can be identified with the transition probabilities of the extended illness-death model which are fully defined by the cause-specific hazard rates (\cite{andersen1993statistical}). 

We considered constant cause-specific hazards and time-varying Weibull hazards. The R code for the simulation study was based on the code by \cite{heggland2015estimating}. The sample sizes were 2,000 and 10,000 observations. Both simulation scenarios have been run 100 times. For both scenarios and each run, we obtained separate and smoothed estimates of both $PAF_{LM,h}$ and $PAF_{0 LM,h}$. Estimation of $PAF_{LM,h}$ was performed as described in Section 3. We did not model any confounding in the simulation study. Therefore, estimation of $PAF_{0 LM,h}$ was based on the unadjusted pseudo-value approach explained in Section 4. Smoothed estimates of $PAF_{LM,h}$ and $PAF_{0 LM,h}$ were obtained via the supermodel approach.
The results of each scenario were summarized as plots of the mean, median, and the first and third quartiles of the 100 runs at each LM.

\begin{figure*}[hbtp!]
\centering
\includegraphics[width=\textwidth]{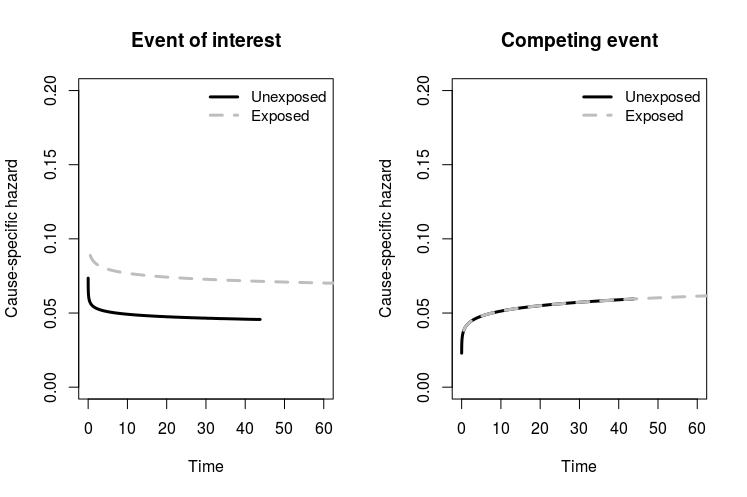}
\caption{Cause-specific Weibull hazard rates with a direct effect of exposure due to an increased hazard of the event of interest with exposure. The hazard of exposure was $\alpha_{01}=0.06$ and time constant.}
\label{fig:haz13}
\end{figure*}

First, we considered a no-effects model meaning, that the exposure does not increase the risk of the event of interest. The no-effects model was simulated with constant hazards. Second, the effect of the exposure was modelled such that it increased the risk of the event of interest directly via an increased hazard of the event of interest. This scenario was based on time-varying cause-specific hazards. The Weibull hazards of this scenario are shown in Figure \ref{fig:haz13}.

The choice of LMs and the time window were based on the data situation. The LMs were chosen such that there were at least 20 observations of exposed and unexposed patients at each LM. The time window was chosen as the mean time at risk. For the time constant-hazards data setting this was 30 time points. In the time-varying-hazards data setting events occurred earlier and the mean duration at risk was eight time points. 

\begin{figure*}[hbtp!]
\centering
\includegraphics[width=\textwidth]{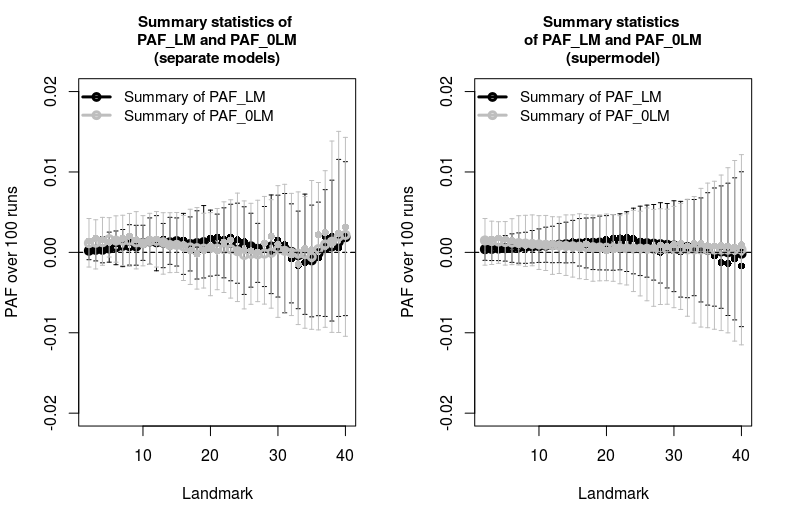}
\caption{Simulation of a no-effects extended illness-death model with constant hazards ($\alpha_{01}(t)=0.005$, $\alpha_{02}(t)=0.02$, $\alpha_{03}(t)=0.02$, $\alpha_{14}(t)=0.02$, and $\alpha_{15}(t)=0.02$). Each sample consists of 10,000 observations. 
The time window was the approximate mean time at risk (30 time points). The summary comprises the mean, median and the first and third quartiles of 100 runs at each LM.}
\label{fig:SimScene1}
\end{figure*}

\begin{figure*}[hbtp!]
\centering
\includegraphics[width=\textwidth]{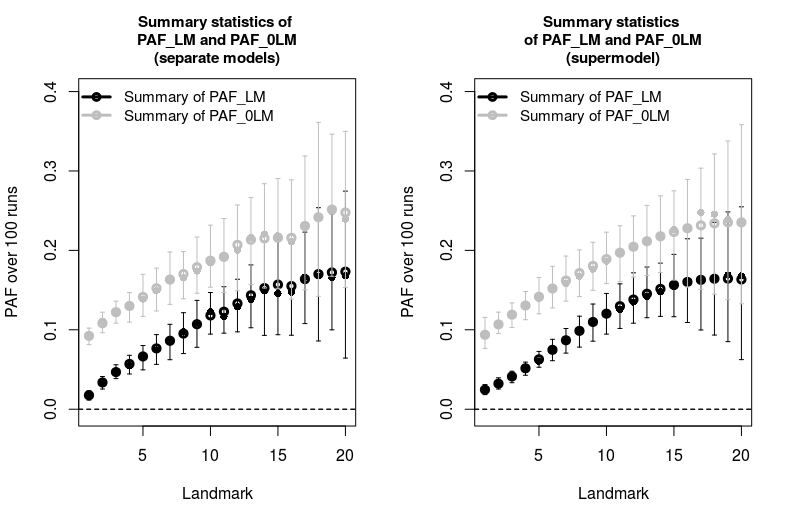}
\caption{Simulation of an extended illness-death model with direct effect of the exposure on the risk to experience the event of interest with time-varying cause-specific Weibull hazard rates (the cause-specific hazards are shown in Figure \ref{fig:haz13}). Each sample consists of 2,000 observations. 
The time window was the approximate mean time at risk (8 time points). The summary comprises the mean, median and the first and third quartiles of 100 runs at each LM. }
\label{fig:SimScene13}
\end{figure*}

In the no-effects model both $PAF_{0 LM,h}$ and $PAF_{LM,h}$ were approximately zero (Figure \ref{fig:SimScene1}).
In the other scenario, $PAF_{0 LM,h}$ was larger than $PAF_{LM,h}$ (Figure \ref{fig:SimScene13}). If exposure could be prevented over the complete considered time window and exposure increases the risk of the event of interest, then at least as many cases are preventable as in a setting where exposure can be only prevented at the LM.

Performance of the estimators varied with sample size at the LMs. Compared to the full cohort there is a significant reduction of patients being exposed and unexposed at the LMs and the number of events occurring in each group. Therefore, variation was considerably large (see Figure \ref{fig:SimScene1} and \ref{fig:SimScene13}). 

The smoothed versions of $\widehat{PAF}_{LM,h}$ and $\widehat{PAF}_{0 LM,h}$ did reduce the variability of the separate estimates, but not considerably. Despite the smoother presentation of the estimates, the simulation study did not show a benefit of the supermodel that would justify the more complex estimation procedure. However, we emphasize that our simulation study does not allow for a generalization of these findings. In Section 6, we further discuss the supermodel in the context of the data example. Source code to reproduce the results is available as Supporting Information on the journal’s web page.

\section{Data example: Preventable death cases among ventilated patients in intensive care}

The prevention of hospital-acquired infections caused by multi-drug resistant pathogens is of major interest to public health. In the following, we investigate the OUTCOMEREA French multi-center database. We consider a sample of 7221 invasive-mechanically ventilated (IMV) patients to understand the burden of ventilator-associated pneumonia caused by the pathogen \textit{Pseudomonas aeruginosa} ($VAP_{P.a.}$) on a population level. Patients are considered at risk of $VAP_{P.a.}$ acquisition after two days of IMV. Follow-up was from first day of IMV to death in the ICU (n=1971) or discharge alive from the ICU (n=5250). Follow-up was complete. However, discharge alive must be considered as competing risk to death in the ICU. Moreover, all patients who enter the study are initially unexposed, $VAP_{P.a.}$ is a time-dependent exposure. By the end of follow-up, 463 patients had acquired a $VAP_{P.a.}$. 

In a previously published competing risks analysis we found that patients with a $VAP_{P.a.}$ have an increased risk of death in the ICU due to a prolonged length of stay (\cite{vonCube2018relative}). Thus, while there was no direct effect on the death hazard (adjusted hazard ratio (adjHR)=1.05; 95\%CI [0.88;1.26]), the hazard of discharge alive was significantly reduced for an infected patient (adjHR=0.67; 95\%CI [0.50;0.79]) (see \cite{vonCube2018relative}).

To understand the public health impact of $VAP_{P.a.}$ for mechanically ventilated patients in intensive care, we estimated both $PAF_{LM,h}$ and $PAF_{0 LM,h}$. The two approaches allow for a differentiation between 'short-stayers' and 'long-stayers'. While most patients leave the ICU after a few days, some patients stay for quite some time. These long-stayers are at higher risk of $VAP_{P.a.}$ acquisition. At the same time, patients are most vulnerable in the first days of IMV. Our approach reveals which patient population would benefit most from a preventive intervention.

\begin{figure*}[hbtp!]
\centering
\includegraphics[width=\textwidth]{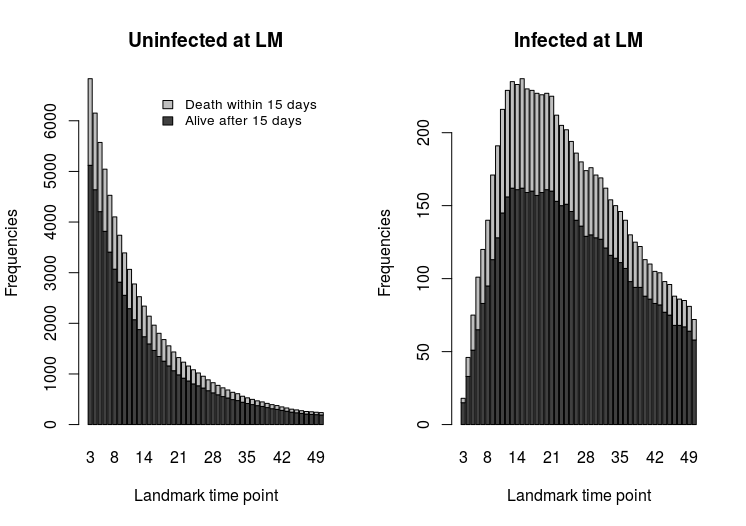}
\caption{Total number of unexposed and exposed patients at each LM of our sample (n=7221) of the OUTCOMEREA database, and the number of ICU death cases within 15 days in each group.}
\label{fig:OUT_LMrisksets}
\end{figure*}

The LMs were chosen daily from day three to day fifty, which is the time frame in which a considerable number of patients were still at risk. The time window was the mean length of stay in the ICU which is 15 days. Figure \ref{fig:OUT_LMrisksets} shows the number of patients at risk among the unexposed and exposed patients at each LM, as well as the number of events occurring within the time window.

We estimated both unadjusted and adjusted versions of $PAF_{LM,h}$ and $PAF_{0 LM,h}$. To account for major differences between infected and uninfected patients, we adjusted for the baseline confounding factors 'type of patient' (surgical versus medical) and age at study entry. Moreover, we adjusted for the time-varying sepsis-related organ failure assessment (SOFA) score, which is an indicator for the patients' severity of the initial illness. To estimate $PAF_{LM,h}$, the SOFA score was updated at each LM and held fixed over the time window. In contrast, when estimating $PAF_{0 LM,h}$, time-varying confounding due to the variation of the SOFA score within the time window must be accounted for. Therefore, we adjusted $PAF_{0 LM,h}$ for the time-varying SOFA score. To further avoid collider-stratification bias we used the values lagged by two days. Estimation of $PAF_{LM,h}$ was performed as described in Section 3. To estimate $PAF_{0 LM,h}$ we accounted for time-varying confounding within the time window. Therefore, we used the inverse- probability weights as explained in Section 4.

To demonstrate the performance of the supermodel on the data example, we obtained smoothed estimators of $PAF_{LM,h}$ using polynomials of degree two as smoothing functions. Thus, we included interaction terms for the baseline risk and for the infection risk. The resulting model is the one presented in Section 3.

\begin{figure*}[hbtp!]
\centering
\includegraphics[width=\textwidth]{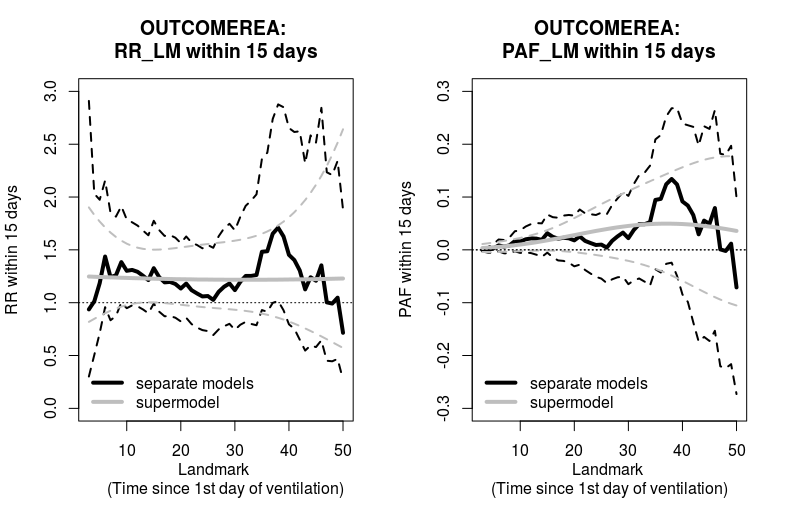}
\caption{Unadjusted relative risk of death (left panel) and population-attributable fraction (right panel) within 15 days depending on the infection state \textit{at} the LM. The dashed black lines are the pointwise 95\%-CI intervals of the separate models (separate, non smoothed estimates). The dashed grey lines are the bootstrap CIs of the supermodels (smoothed estimates).}
\label{fig:OUT_RR_PAFlm}
\end{figure*}

To derive $PAF_{LM,h}$ we also estimated $RR_{l,h}$. The unadjusted $RR_{l,h}$ (for all LMs $l$, $l\in \mathcal{LM}$) and $PAF_{LM,h}$ (separate models and supermodel) are shown in Figure \ref{fig:OUT_RR_PAFlm}. We observe an increased RR of death for patients still in the ICU at day 38 ($\widehat{RR}_{38,15}=1.7$, 95\%-CI $[1.02; 2.88]$, separate models). Due to this increased RR the PAF (separate model) also has a peak at day 38 ($\widehat{PAF}(38,15)=0.13$, 95\%-CI $[-0.024; 0.27]$). The zero is contained in the 95\%-CI. This implies that there is no statistical evidence for attributable mortality of $VAP_{P.a.}$ within 15 days for patients in the ICU at day 38. In general, there is no evidence for attributable mortality within 15 days at any of the LMs, if the reference population are those patients that are unexposed at the LM.

The supermodel for both RR and $PAF_{LM,h}$ clearly increases the efficiency of the estimator. The point-wise CIs are considerably smaller. Nevertheless, the supermodel has some drawbacks. The supermodel puts more weight on LMs with many patients at risk. As the sample size is decreasing with time, more weight is put on early LMs. The separate models of $\widehat{RR}_{LM,h}$ indicate two peaks at day six and day 38. Due to the smaller sample size at day 38, the supermodel does not follow this trend. The data example demonstrates that the supermodel should be interpreted in combination with the separate models.

\begin{figure*}[hbtp!]
\centering
\includegraphics[width=\textwidth]{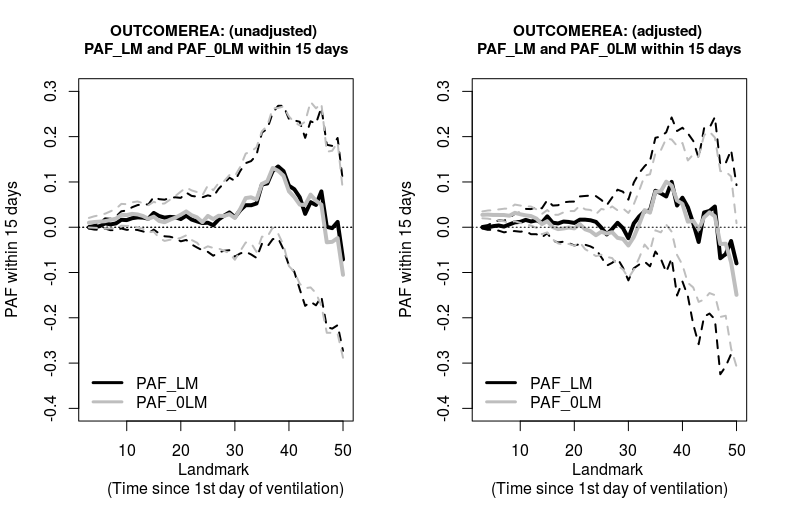}
\caption{Unadjusted estimates of $PAF_{LM,h}$ and $PAF_{0 LM,h}$ (left panel) and the adjusted estimates (right panel) of our sample of the OUTCOMEREA database. The dashed lines are the pointwise 95\%-CI intervals of the separate models of $PAF_{LM,h}$ (black) and $PAF_{0LM,h}$ (grey).}
\label{fig:OUT_lmVSlm0}
\end{figure*}

The left panel of Figure \ref{fig:OUT_lmVSlm0} shows the unadjusted estimates of $PAF_{LM,h}$ and $PAF_{0 LM,h}$ (separate models). $\widehat{PAF}_{0 LM,h}$ indicates attributable mortality within 15 days at LMs nine ($\widehat{PAF}_0(9,15)=0.028$, 95\%-CI $[0.004; 0.05]$), ten ($\widehat{PAF}_0(10,15)=0.027$, 95\%-CI $[0.002; 0.05]$), and eleven ($\widehat{PAF}_0(11,15)=0.03$, 95\%-CI $[0.002; 0.06]$) if the reference population are those patients that are unexposed within $[l,l+h]$. For example, if $VAP_{P.a.}$ could be avoided between days 11 to 26 for patients still in the ICU at day 11, then 3\% of the death cases occurring within this time window could be prevented. This corresponds to a total number of 25 cases. 
Nevertheless, as the prevalence of $VAP_{P.a.}$ is -- from a statistical point of view -- very low, there is no remarkable difference between $\widehat{PAF}_{LM,h}$ and $\widehat{PAF}_{0 LM,h}$.

The adjusted versions of $\widehat{PAF}_{LM,h}$ and $\widehat{PAF}_{0 LM,h}$ are shown in the right panel of Figure \ref{fig:OUT_lmVSlm0}. The estimates reveal that the patient populations are most vulnerable within the first days of IMV (days three to twelve). Here targeted preventive interventions against $VAP_{P.a.}$ with a long term effect of 15 days would provide the most benefit. A plot that contrasts the adjusted versions of $\widehat{PAF}_{LM,h}$ and $\widehat{PAF}_{0 LM,h}$ with the unadjusted ones is provided in the online supplementary material in Figure S2. Source code to reproduce the results is available as Supporting Information on the journal’s web page.

\section{Data example: TEAM trial data}

\cite{fontein2015dynamic} studied the risk of death for a sample of the Tamoxifen Exemestane Adjuvant Multinational (TEAM) trial. The randomized clinical trial included Belgian and Dutch early breast cancer (BC) patients that were postmenopausal hormone receptor-positive (HR+) and treated with endocrine. Fontein et al. used dynamic prediction and landmarking (\cite{van2011dynamic, fontein2015dynamic}) for survival prognosis being made months past time of diagnosis. The approach accounts for time-varying effects and allows for an update of patients characteristics recorded past baseline. \cite{fontein2015dynamic} found a highly significant increase in all-cause mortality for patients experiencing locoregional recurrence (LRR) or distant recurrence (DR).

We extended the analysis to study the population-attributable burden of LRR and DR. The outcome of interest was death of any cause. Thus, in contrast to the data example in Section 6, we have no competing risks. However, of the 2597 patients included in the analysis, 2238 were censored due to end of follow-up.
Of the initially included patients, 88 eventually experienced LRR and 406 DR (48 had both LRR and DR, see also the online supplementary material). A precise clinical definition of LRR and DR is provided by \cite{fontein2015dynamic}. Both DR and LRR are time-varying exposures. A multi-state model illustrating the data setting is provided in the online supplementary material in Figure S3. 

To estimate $PAF_{LM,h}$ and $PAF_{0LM,h}$, we used a time window of five years. The LMs were chosen at every third month from the second year until the fourth year since diagnosis. In contrast to \cite{fontein2015dynamic}, we do not consider any LMs before the first year. Since we use the approach 'dynamic prediction by landmarking' for inference rather than prediction a sufficient number of exposed patients should be present at the LMs. The number of unexposed and exposed patients at each LM for both exposures (LRR and DR) as well as the number of death cases within the time window of 5 years is illustrated in bar plots in the online supplementary material in Figure S4.

We obtained unadjusted and adjusted estimates of $PAF_{LM,h}$ and $PAF_{0 LM,h}$ for both exposures, DR and LRR. To demonstrate different ways of estimation, estimation of $PAF_{LM,h}$ was based on definition \eqref{PAF(s,t)_lev} rather than definition \eqref{PAF(s,t)_def} which has been used in the previous data example. The observable overall mortality risk at the LMs within the time window was the same for both estimates, $\widehat{PAF}_{LM,h}$ and $\widehat{PAF}_{0 LM,h}$, and both exposures. Regarding $\widehat{PAF}_{LM,h}$, estimation of the death risk among unexposed was based on the patients being unexposed at the LM. Regarding $\widehat{PAF}_{0 LM,h}$, the estimate was obtained by treating patients experiencing LRR or respectively DR within the time window of five years as administratively censored at the time of recurrence. To obtain adjusted estimates, we adjusted the mortality risk of unexposed patients for each estimand ($PAF_{LM,h}$ and $PAF_{0 LM,h}$) with the Cox proportional dynamic prediction model as proposed by \cite{fontein2015dynamic}. A more detailed explanation is provided in the online supplementary material.

The covariates included in the model were age at diagnosis (as continuous variable, constant and squared), Bloom \& Richardson (BR) histological grade (I, II, III), tumor stage (1,2, 3/4), nodal stage (N0, N1, N2/N3), ER and PR status (positive, negative), HER2 status (positive, negative, missing), most extensive surgery (mastectomy, breast conserving surgery), radiotherapy (yes, no). chemotherapy (yes/no) and treatment status (on/off). Moreover, we adjusted both $PAF_{LM,h}$ and $PAF_{0 LM,h}$ of DR for LRR status at the LM and -- as Fontein et al. -- included an interaction term for LRR and LM (\cite{fontein2015dynamic}). Similarly, we adjusted $PAF_{LM,h}$ and $PAF_{0 LM,h}$ of LRR for DR status at the LM.
A detailed description of the covariates, as well as the regression coefficients of the dynamic Cox model, can be found in the article by \cite{fontein2015dynamic}.

\begin{figure*}[hbt!]
\centering
\includegraphics[width=\textwidth]{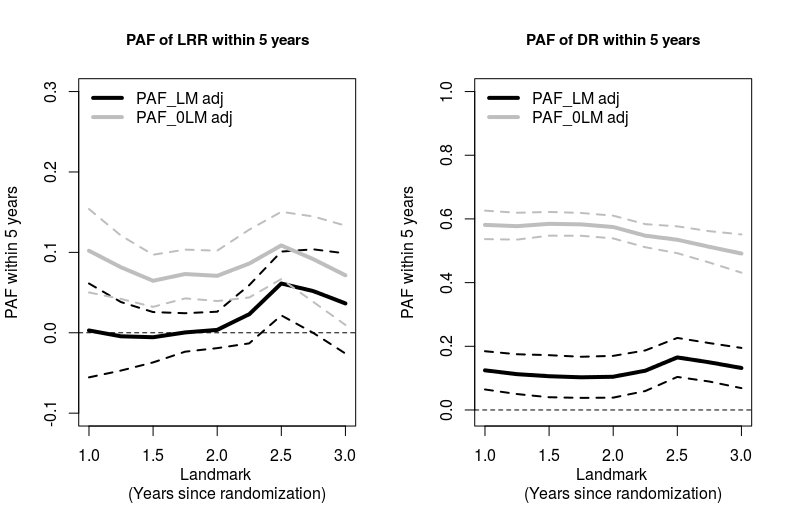}
\caption{Adjusted estimates of $PAF_{LM,h}$ and $PAF_{0 LM,h}$ of LRR (left panel) and DR (right panel) of the TEAM trial data sample. The dashed lines are the pointwise 95\%-CI intervals of the separate models of $PAF_{LM,h}$ (black) and  $PAF_{0LM,h}$ (grey).}
\label{fig:teamPAF}
\end{figure*}

The adjusted estimates of both $PAF_{LM,h}$ and $PAF_{0 LM,h}$ of LRR and DR are shown in Figure \ref{fig:teamPAF}. 
A plot that contrasts the adjusted versions of $\widehat{PAF}_{LM,h}$ and $\widehat{PAF}_{0 LM,h}$ with the unadjusted ones is provided in the online supplementary material in Figure S5. First, we consider $\widehat{PAF}_{LM,h}$ and $\widehat{PAF}_{0 LM,h}$ of LRR, which are shown in the left panel of Figure \ref{fig:teamPAF}. The estimate $\widehat{PAF}_{LM,h}$ is approximately zero at all LMs but reaches a peak at 2.5 years post randomization. At this LM only the CI becomes greater than 0. In contrast, an intervention that could prevent LRR for 5 years would be beneficial at all LMs.

Next, we consider $\widehat{PAF}_{LM,h}$ and $\widehat{PAF}_{0 LM,h}$ of DR, which are shown in the right panel of Figure \ref{fig:teamPAF}. Both types of intervention would be beneficial at all LMs, with a clear advantage of an intervention that is effective over the next five years. Then, for example at LM 1 almost 60\% (adj. $\widehat{PAF}_{0 LM,h} \approx 0.55$, 95\%-CI $\approx [0.5; 0.6]$) of the death cases occurring within five years since the LM would be preventable. In comparison, only about 12\% (adj. $\widehat{PAF}_{LM,h} \approx 0.12$, 95\%-CI $\approx [0.06; 0.18]$) would be preventable when the intervention is effective at the LM only.

With the data example, we demonstrated the application of the proposed approach for a data setting with a binary time-dependent exposure in a basic survival setting (no competing risks). To obtain adjusted estimates we used adjusted survival function based on the Cox proportional hazards model (see also the online supplementary material). Source code to reproduce the results is available as Supporting Information on the journal’s web page.

\section{Discussion}

This paper introduces a novel approach to define and estimate the PAF for complex data situations. Time-dependent exposures are common in epidemiology. However, statistical modelling of the time dynamics of exposure and outcome is challenging and often avoided. Our proposed approach addresses these challenges in a very basic manner. Nevertheless, it accommodates complex study designs such as survival data with time-dependent exposure and subject to competing risks and censoring.

The two proposed estimands, namely $PAF_{LM,h}$ and $PAF_{0 LM,h}$, have a clinically relevant interpretation as they account for the dynamics of the population over the course time. 
At a specific LM, $PAF_{LM,h}$ is interpretable as the proportion of preventable cases within a predefined time window if the exposure could be prevented at the LM. Thus, the estimand is based on a hypothetical intervention that would be effective at time of intervention, i.e. at the considered LM, only. In contrast, $PAF_{0 LM,h}$, at a specific LM, is interpretable as proportion of preventable cases within the time window if the exposure could be prevented over the complete time window. The assumed intervention has a long term effect over the complete time window. 

Definition of both estimands was based on dynamic prediction and landmarking (\cite{van2008dynamic}) and estimation was based on existing methods. Adjustment of $PAF_{LM,h}$ for time-dependent covariates is straightforward since at each LM the time-dependent data setting is reduced to a data setting with baseline exposure and fixed length of follow-up. Thus, $PAF_{LM,h}$ retains the advantages of conventional dynamic prediction by landmarking. Since the exposure state is accessed at the LM only, adjustment for confounding can be performed by updating the time-dependent patient characteristics at the LM and including these updated values as time-independent covariates in the model. In contrast, estimation of $PAF_{0 LM,h}$ requires adjustment for time-varying confounding. The time-varying confounding occurs as individuals must be comparable within the complete time window. When estimating $PAF_{LM,h}$ individuals must be comparable at the LM only. As a consequence estimation of $PAF_{0LM,h}$ is more complicated.

Nevertheless, $PAF_{0LM,h}$ relaxes the assumption that the exposure can be prevented at the LM only. Moreover, it overcomes the limitation of data samples in which many individuals acquire the exposure at later LMs. In the conventional approach, used to define $PAF_{LM,h}$, these individuals are considered unexposed at the early LMs. If the exposure is harmful, the individuals who acquire exposure within the time window increase the risk among unexposed and therefore potentially preclude the burden of the exposure at earlier LMs. Our data examples in Section 6 and 7 demonstrated that this effect is negligible if the prevalence is low (data example in Section 6) and strong if the prevalence is high (data example in Section 7).

Originally, dynamic prediction by landmarking has been proposed to make survival prognoses (\cite{van2008dynamic}). Similar to \cite{gran2010sequential}, we applied the approach for inference instead. While unproblematic when making predictions, conditioning on survival up to a certain LM may cause selection bias if the goal is to understand the causal relationship of exposure and outcome (\cite{aalen2015does}). This means that the population at risk at LM $l$ is not a representative sample of the initial study population sampled at baseline. The difference is due to (unmeasurable) confounding that has an effect on survival. However, since our target population are not all initially sampled patients but only those still at risk at LM $l$ selection bias is naturally avoided (\cite{gran2010sequential}).

A remaining drawback is the fact that the two landmark estimands were identified with the assumption that the time from exposure to the LM has no impact on the risk of experiencing the event of interest within the prediction time window. This assumption corresponds to the well-known Markov assumption (\cite{beyersmann2011competing}, \cite{aalen2008survival}). If this assumption cannot be made, we must make stronger assumptions about the hypothetical intervention. Then, we must assume that it prevents the exposure exactly at the LM (and within the time window when considering $PAF_{0LM,h}$), but also reverses/cures any harm caused by the exposure for individuals who were exposed before the LM. At the same time the intervention must be specific in the sense that it reverses \textit{only} effects caused by the exposure. If the individuals are healthy besides the exposure-associated health issues such an intervention seems plausible. 
However, in data settings where individuals are critically ill, as for example in an ICU setting, such a specific intervention seems rather unrealistic. Despite these interpretational limitations in these data settings, the approach provides information about the best timing of an intervention and identifies the most vulnerable target populations.

To demonstrate the interpretation of the estimands and performance of the estimators, we provided a simulation study and two real data examples. Firstly, we investigated the benefit of a pathogen-specific intervention against $VAP_{P.a.}$. Secondly, we investigated the population-attributable burden of LRR and DR in a population of breast cancer patients. The two data examples served to explain the interpretation of the novel approach. We based our interpretation of $PAF_{LM,h}$ and $PAF_{0LM,h}$ on two distinct hypothetical interventions. The way the interventions act on exposure may be more or less realistic depending on the data setting. For example, an intervention that prevents LRR or DR at only one specific time point is less realistic than an intervention with a long term effect. In these situations, the estimates can be used to identify high-risk populations and to describe the burden of exposure on a population level.

A disadvantage of the proposed estimation procedure is the sample size reduction at the LMs, which leads to a comparatively large variation of $\widehat{PAF}_{0 LM,h}$ and $\widehat{PAF}_{LM,h}$. The smoothing methods currently available may either result in a loss of information or a negligible increase in efficiency.

To sum up, we proposed two novel estimands of the PAF to quantify the burden of time-dependent harmful exposures on a population level. The two estimands also account for competing risks, filling a gap in the literature (\cite{sjolander2014doubly}). Finally, our innovative way of using dynamic prediction and landmarking allows for adjustment of time-varying confounding in a straightforward way.

\newpage

\section*{Figures}

\clearpage

\textbf{Acknowledgements}
The authors thank Mia Klinten Grand, postdoctoral researcher at the Department of Public Health, University of Copenhagen, for providing helpful information and R code for the analysis of the TEAM trial data example.

\vspace*{1pc}

\noindent {\bf{Conflict of Interest}}

\noindent {\it{The authors have declared no conflict of interest.}}

\noindent {\bf{Funding}}
\noindent {MvC and JFT were supported by the Innovative Medicines Initiative Joint Undertaking under grant agreement n [115737-2 – COMBACTE-MAGNET], resources of which are composed of financial contribution from the European Union’s Seventh Framework Programme (FP7/2007-2013) and EFPIA companies; MW has received funding from the German Research Foundation (Deutsche Forschungsgemeinschaft) under grant no. WO 1746/1-2.}

{
\bibliographystyle{ieeetr}
\bibliography{databaseCOMBACTEMagnet}
}

\end{document}